\documentstyle[aps, epsf, psfig]{revtex}
\begin{document}
\draft
\title{CP violation in $\psi(2S)\rightarrow J/\psi \pi \pi$ processes}
\author{Jie-Jie Zhu$^{1,2}$\footnote{Electronic address: jjzhu@ustc.edu.cn},
Mu-Lin Yan$^2$}
\address{
$^1$CCAST(World Lab), P. O. Box 8730, Beijing, 100080, P. R.
China\\
$^2$Center for Fundamental Physics, USTC,
Hefei, Anhui, 230026, P. R. China\footnote{Permanent address.} }
\date{\today}
\maketitle
\begin{abstract}
We propose to search for CP-violating effects in the decay
$\psi(2S)\rightarrow J/\psi \pi \pi$. The scheme has the advantage
that one does not need to track two or more CP-conjugate processes.
Model independent amplitudes are derived for this purpose. The fact
that leading CP violating terms are ${\cal O}(k)$ under low energy
expansion and the processes are flavor disconnected
make the measurement of these CP breaking parameters
practical. Our results can be extended to the case of
$\Upsilon(2S)\rightarrow\Upsilon(1S) \pi \pi$ and
$\Upsilon(3S)\rightarrow\Upsilon(2S) \pi \pi$ straightforwardly.
\end{abstract}

CP violation is a subject attracting much interest. In the Standard
Model, it arises as a phase entering the CKM matrix. It is believed
that, with the CP violation presented in the Standard Model, it is
not possible to generate the observed size of matter-antimatter
asymmetry of the Universe~\cite{cp99,sakharov67}. 
However, CP violation has only
been observed in neutral-kaon systems till now. The evidence come
from the measurements of $\eta_{+-}$, $\eta_{00}$, and the
semileptonic decay charge asymmetry for $K_L$. Currently
experimental efforts are concentrated on neutral systems such as
$K^0$-$\bar{K}^0$, $B^0$-$\bar{B}^0$, and $D^0$-$\bar{D}^0$
complex. Other searches, as summarized by Wolfenstein and Trippe,
devide into two classes: (a) Those that involve weak interactions
or parity violation. The most sensitive are the searches for an
electric dipole moment of the elementary particles such as neutron
and electron. (b) Those that involve processes otherwise allowed by
the strong or electromagnetic interactions. This includes the
search for C violation in $\eta$ decay and searches for T violation
in a number of nuclear and electromagnetic reactions~\cite{pdg98}.

In this paper we suggest a new means to search for CP violation. We
propose to measure CP asymmetries in the decay $\psi(2S)\rightarrow
J/\psi\pi^+\pi^-$ or $\psi(2S)\rightarrow J/\psi\pi^0 \pi^0$. 
The process $\psi(2S)\rightarrow J/\psi\pi^+\pi^-$ has been discussed 
based on an effective lagrangian~\cite{mannel97,yan99},  
the multipole expansion hypothesis~\cite{voloshin80,novikov81,yan80}, 
or current algebra~\cite{brown75}. 
These discussions are model dependent, and C, P, T invariance are presumed. 
We derive  model independent amplitudes of the two processes.
CP invariance requires particular parameters in the decay amplitude vanish.
Any non-zero value of those particular parameters implies CP violation. 
One will not need to track two or more channels at the same time. 
And we need not to compare the phases or decay rate of two or more 
CP-conjugate processes to get CP violation observables, thus avoid precision
loss when we subtracting two very close numbers or information loss
because of using total transition rates instead of differential cross
sections.

$J/\psi\pi^+ \pi^-$ and $J/\psi\pi^0 \pi^0$ are the two largest
decay modes of $\psi(2S)$, with branching ratios $(30.2 \pm 1.9)\%$
and $(17.9 \pm 1.8)\%$~\cite{pdg98}. The masses for $\psi(2S)$ and $J/\psi$ are
$3686.00\pm 0.09$ MeV and $3096.88\pm 0.04$ MeV, with a difference 
of 589 MeV. The energy available for the
pions are small and as we will shown, most part of the amplitude
are expected to be the contribution of contact interactions. This
validates the low energy expansion. The leading CP-violating terms
in the model independent amplitudes are of the first order of the
soft pion momentum(${\cal O}(k)$). BES has enough data to see
D-wave contributions~\cite{bes99}, which is ${\cal O}(k^2)$. 
Because the processes are flavor disconnected, the strong interactions
are suppressed, and we have chances to see CP-violating effects
beyond (or even within) the Standard Model in these decays.
So it is practical to measure these CP-violating parameters. 
It will not take long for BEPC to accumulate 
$10^8$ $\psi(2S)$ events. Now BEPC is scheduled for another
upgrade in the near future. After the upgrade it will have the
ability to take more than $10^9$ $\psi(2S)$ events and give a high 
precision test of CP invariance in $\psi(2S)\rightarrow J/\psi\pi\pi$ 
decays.

The Feynman amplitude for the process $\psi(2S)\rightarrow J/\psi
\pi^+ \psi^-$ reads
\begin{equation}
{\cal
M}_{\lambda,\sigma}(p,q,k_1,k_2)=e^\mu_{\psi'}(\vec{p},\lambda)
e^{*\nu}_{J/\psi}(\vec{q},\sigma) \Gamma_{\mu\nu}.
\end{equation}
Here $p, q, k_1, k_2$ are four-momenta of the particles $\psi(2S),
J/\psi, \pi^+, \pi^-$. The helicities of $\psi(2S)$ and $J/\psi$
are $\lambda$ and $\sigma$, while $e^\mu_{\psi'}(\vec{p},\lambda)$
and $e^{*\nu}_{J/\psi}(\vec{q},\sigma)$ are the corresponding
polarization vectors for the two particles. $\Gamma_{\mu\nu}$ is
the effective vertex for the process, which contains all details
of the decay.

\vspace{0.5cm}
\begin{figure}[htb]
\centerline{\psfig{figure=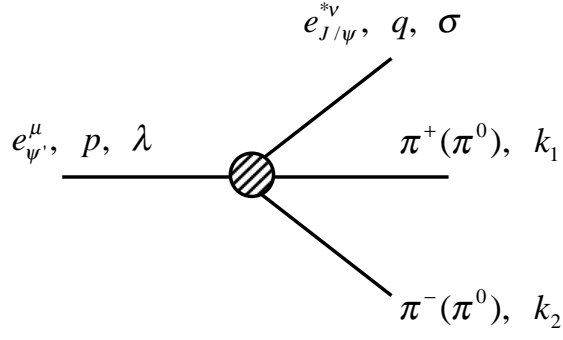,width=3in}}
\caption{$\psi(2S)$ decays into $J/\psi$ and two pions.}\label{fig1}
\end{figure}

The effective vertex $\Gamma^{\mu\nu}$ should be constructed from
the four-momenta $p^\alpha$, $q^\alpha$, $k_1^\alpha$, $k_2^\alpha$
and isotropic tensors $g^{\alpha\beta}$,
$\varepsilon^{\alpha\beta\gamma\delta}$. The effective vertex
contains only the zeroth and first order of the antisymmetric
tensor because of the identity
\begin{equation}
\varepsilon_{\alpha \beta \gamma \delta}
        \varepsilon_{\alpha^{'} \beta^{'} \gamma^{'} \delta^{'}}
= -\det
\left(\begin{array}{cccc} g_{\alpha \alpha^{'}} & g_{\beta
 \alpha^{'}}&g_{\gamma \alpha^{'}} & g_{\delta \alpha^{'}} \\
g_{\alpha \beta^{'}} & g_{\beta \beta^{'}} &
        g_{\gamma \beta^{'}} & g_{\delta \beta^{'}} \\
g_{\alpha \gamma^{'}} & g_{\beta \gamma^{'}} &
         g_{\gamma \gamma^{'}} & g_{\delta \gamma^{'}} \\
g_{\alpha \delta^{'}} & g_{\beta \delta^{'}} &
        g_{\gamma \delta^{'}} & g_{\delta \delta^{'}}
\end{array}\right).
\end{equation}
Considering energy-momentum conservation
\begin{eqnarray}
p^\alpha = q^\alpha + k_1^\alpha + k_2^\alpha
\end{eqnarray}
and the Lorentz conditions for polarization vectors
\begin{eqnarray}
p_\alpha e^\alpha_{\psi'}(\vec{p},\lambda) &=& 0,
\\ q_\alpha
e^{*\alpha}_{J/\psi}(\vec{q},\sigma) &=& 0,
\end{eqnarray}
we find the general form of the effective vertex
\begin{eqnarray}
& &
\Gamma^{\mu\nu}(p,q,k_1,k_2,g^{\alpha\beta},\varepsilon^{\alpha\beta\gamma\delta}) \nonumber \\
&=& c_1g^{\mu\nu} +c_2k_1^\mu k_1^\nu +c_3k_2^\mu k_2^\nu
+c_4k_1^\mu k_2^\nu +c_5k_2^\mu k_1^\nu +c_6A_1^{\mu\nu}
+c_7A_2^{\mu\nu}
\nonumber \\ & & +c_8 A_3^{\mu\nu} +c_9 Q^\mu k_1^\nu +c_{10}Q^\mu k_2^\nu
+c_{11}k_1^\mu Q^\nu +c_{12}k_2^\mu Q^\nu,
\end{eqnarray}
where we define
\begin{eqnarray}
A_1^{\alpha\beta} &=& p_\gamma k_{1\delta}
\varepsilon^{\alpha\beta\gamma\delta},\\
A_2^{\alpha\beta} &=& p_\gamma k_{2\delta}
\varepsilon^{\alpha\beta\gamma\delta},\\
A_3^{\alpha\beta} &=& k_{1\gamma} k_{2\delta}
\varepsilon^{\alpha\beta\gamma\delta},\\
Q^\alpha &=& p_\beta k_{1\gamma} k_{2\delta}
\varepsilon^{\alpha\beta\gamma\delta}.
\end{eqnarray}

The twelve form factors in $\Gamma^{\mu\nu}$ are not independent
since
\begin{eqnarray}
Q_\mu k_{1\nu} e^\mu_{\psi'}e^{*\nu}_{J/\psi} \equiv
\left(k_{1\mu} Q_\nu +(k_1\cdot k_2)A_{1\mu\nu} -k_1^2A_{2\mu\nu} +(p\cdot k_1)A_{3\mu\nu}
\right)e^\mu_{\psi'}e^{*\nu}_{J/\psi},\\
Q_\mu k_{2\nu} e^\mu_{\psi'}e^{*\nu}_{J/\psi} \equiv
\left(k_{2\mu} Q_\nu +k_2^2A_{1\mu\nu} -(k_1\cdot k_2)A_{2\mu\nu} +(p\cdot k_2)A_{3\mu\nu}
\right)e^\mu_{\psi'}e^{*\nu}_{J/\psi},\\
\left(Q_\mu k_{1\nu} +Q_\mu k_{2\nu}
\right)e^\mu_{\psi'}e^{*\nu}_{J/\psi} \equiv
\left(
(p\cdot k_2)A_{1\mu\nu} -(p\cdot k_1)A_{2\mu\nu} +p^2A_{3\mu\nu}
\right)e^\mu_{\psi'}e^{*\nu}_{J/\psi}.
\end{eqnarray}
We can eliminate three terms in the effective vertex without
introducing kinematic singularities. After a redefinition of form
factors we have nine independent terms,
\begin{eqnarray}\label{eq:id}
 & &
\Gamma^{\mu\nu}(p,q,k_1,k_2,g^{\alpha\beta},\varepsilon^{\alpha\beta\gamma\delta})\nonumber \\
&=& c_1g^{\mu\nu} +c_2\left(k_1^\mu k_1^\nu +k_2^\mu k_2^\nu\right)
+c_3\left(k_1^\mu k_1^\nu -k_2^\mu k_2^\nu\right) +c_4\left(k_1^\mu
k_2^\nu +k_2^\mu k_1^\nu\right) \nonumber\\ & & +c_5\left(k_1^\mu
k_2^\nu -k_2^\mu k_1^\nu\right)
+c_6\left(A_1^{\mu\nu}+A_2^{\mu\nu}\right)
+c_7\left(A_1^{\mu\nu}-A_2^{\mu\nu}\right) \nonumber \\ & &
+c_8A_3^{\mu\nu} +c_9Q^\mu\left(k_1^\nu-k_2^\nu\right),
\end{eqnarray}
with
\begin{equation}
c_i=c_i\left((k_1\cdot k_2), [p\cdot (k_1-k_2)]\right)
\end{equation}
for $i=1, 2,\cdots, 9$.

Expand the form factors in Taylor series of $(k_1\cdot k_2)$ and
$[p\cdot (k_1-k_2)]$, and keep those terms up to ${\cal O}(k^2)$,
\begin{eqnarray}\label{eq:p+p-}
& &
\Gamma^{\mu\nu}(p,q,k_1,k_2,g^{\alpha\beta},\varepsilon^{\alpha\beta\gamma\delta})
\nonumber \\
&=& \{f_0 + \frac{[p\cdot(k_1-k_2)]}{M_{\psi'}^2}f_1
+\frac{(k_1\cdot k_2)}{M_{\psi'}^2}f_2 +\frac{[p\cdot
(k_1-k_2)]^2}{M_{\psi'}^4}f_3 \}g^{\mu\nu} \nonumber
\\ & & +\frac{(k_1^\mu k_1^\nu +k_2^\mu k_2^\nu)}{M_{\psi'}^2}f_4
+\frac{(k_1^\mu k_1^\nu-k_2^\mu k_2^\nu)}{M_{\psi'}^2}f_5 \nonumber
\\ & & +\frac{(k_1^\mu k_2^\nu +k_2^\mu k_1^\nu)}{M_{\psi'}^2}f_6
+\frac{(k_1^\mu k_2^\nu-k_2^\mu k_1^\nu)}{M_{\psi'}^2}f_7 \nonumber
\\ & & +i\frac{(A_1^{\mu\nu}+A_2^{\mu\nu})}{M_{\psi'}^2}f_8
+i\frac{(A_1^{\mu\nu}+A_2^{\mu\nu})[p\cdot
(k_1-k_2)]}{M_{\psi'}^4}f_9
\nonumber
\\ & & +i\frac{(A_1^{\mu\nu}-A_2^{\mu\nu})}{M_{\psi'}^2}f_{10}
+i\frac{(A_1^{\mu\nu}-A_2^{\mu\nu})[p\cdot
(k_1-k_2)]}{M_{\psi'}^4}f_{11} \nonumber
\\ & & +i\frac{A_3^{\mu\nu}}{M_{\psi'}^2}f_{12},
\end{eqnarray}
where $M_{\psi'}$ is the mass of $\psi(2S)$ and $f_i$ are
dimensionless complex constants. The reason why the energy scale
should be $M_{\psi'}$, and why we extract a factor ``$i$'' from $f_8,
f_9, \cdots, f_{12}$ will be explained later.

Now let's see what CP invariance can say about the form factors. If
CP parity is conserved, then
\begin{eqnarray}
& & \langle \psi'|{\bf S}|J/\psi\pi^+\pi^-\rangle \nonumber \\ & =
& \langle \psi'|{\bf
(CP)^\dagger(CP)S(CP)^{-1}(CP)}|J/\psi\pi^+\pi^-\rangle \nonumber
\\ &=& \langle \psi'|{\bf (CP)^\dagger S(CP)}|J/\psi\pi^+\pi^-\rangle .
\end{eqnarray}
The spin-parity $J^{PC}$ for $\psi(2S)$, $J/\psi$ and pions are
$1^{--}$, $1^{--}$ and $0^{-+}$~\cite{pdg98},
\begin{eqnarray}
{\bf CP} |\vec{p},\lambda\rangle & = &
-|-\vec{p},-\lambda\rangle, \\ {\bf CP} |\vec{q},
\sigma,\vec{k}_1, \vec{k}_2\rangle
& = &
-|-\vec{q},-\sigma,-\vec{k}_2,
-\vec{k}_1\rangle.
\end{eqnarray}
So conservation of CP require
\begin{equation}
{\cal M}_{\lambda,\sigma}(p,q,k_1,k_2) \equiv {\cal
M}_{-\lambda,-\sigma}(\bar{p},\bar{q},\bar{k}_2,\bar{k}_1).
\end{equation}
Where a vector with a bar indicate its space reflected value, e.g.,
\begin{equation}
\bar{q}^\alpha \equiv {\cal P}^\alpha_\beta q^\beta,
\end{equation}
and ${\cal P}$ is the space reflection matrix, $({\cal
P}^\alpha_\beta) = {\rm diag}\{1,-1,-1,-1\}$. That is
\begin{eqnarray}
& & e^\mu_{\psi'}(\vec{p},\lambda)
e^{*\nu}_{J/\psi}(\vec{q},\sigma)
\Gamma_{\mu\nu}(p,q,k_1,k_2,g^{\alpha\beta},\varepsilon^{\alpha\beta\gamma\delta})
\nonumber \\ &=&
e^\mu_{\psi'}(-\vec{p},-\lambda)
e^{*\nu}_{J/\psi}(-\vec{q},-\sigma)
\Gamma_{\mu\nu}(\bar{p},\bar{q},\bar{k}_2,\bar{k}_1,g^{\alpha\beta},
\varepsilon^{\alpha\beta\gamma\delta}) \nonumber \\ &=&
\bar{e}^\mu_{\psi'}(\vec{p},\lambda)
\bar{e}^{*\nu}_{J/\psi}(\vec{q},\sigma)
\Gamma_{\mu\nu}(\bar{p},\bar{q},\bar{k}_2,\bar{k}_1,g^{\alpha\beta},
\varepsilon^{\alpha\beta\gamma\delta}) \nonumber \\
&=& e^\mu_{\psi'}(\vec{p},\lambda)
e^{*\nu}_{J/\psi}(\vec{q},\sigma)
\Gamma_{\mu\nu}(p,q,k_2,k_1,g^{\alpha\beta},-\varepsilon^{\alpha\beta\gamma\delta}).
\end{eqnarray}

Since all form factors in Eq.~(\ref{eq:p+p-}) are independent, 
it is easy to see that
{\em any non-vanishing value of $f_1$,$f_5$,$f_7$,$f_8$ or $f_{11}$
implies CP violation}. $f_1/f_0$, $f_5/f_0$, $f_7/f_0$, $f_8/f_0$
and $f_{11}/f_0$ can be taken as CP breaking parameters. Among them, the
$f_1$ and $f_8$ terms are ${\cal O}(k)$. In fact, the parameters
in Eq.~(\ref{eq:p+p-}) can be classified into four types(as shown in 
Tab.~\ref{tab:par}):
(1)$f_0$, $f_2$, $f_3$, $f_4$ and $f_6$ terms keep both C and
P symmetry. (2)$f_1$, $f_5$ and $f_7$ terms are C nonconserving but keep P
symmetry. (3)$f_8$ and $f_{11}$ terms keep C symmetry but break P.
(4)$f_9$, $f_{10}$ and $f_{12}$ terms break both C and P, but keep CP
invariance.

If we suppose the amplitude is completely the first order
contribution of an effective lagrangian, and the lagrangian is
time reversal invariant, the form factors $f_0, f_2, \cdots,
f_{12}$ will be relatively real. That's why we extract a factor
"$i$" in the abnormal terms. Non-zero relative phases may come from T
violation in the effective lagrangian or high order terms(loops and
resonance). However, since there is no strong resonance in the process, we
can expect the Feynman amplitude to be dominated by contact interaction
terms in the effective lagrangian, so that their relative phases will be
small. At least this will be true for those terms keeping both CP
and isospin symmetry, i.e., the relative phases between
$f_0,f_2,f_3,f_4,f_6$ will be very small(but one should not mistake 
a nonzero relative phase as the signal of CP violation).

\vspace{0.5cm}
\begin{figure}[htb]
\centerline{\psfig{figure=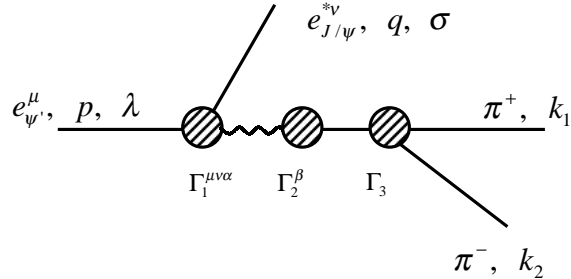,width=3in}}
\caption{Gluonic transition to a spin-0 resonance.
The diagram gives a zero contribution due to gauge
invariance.}\label{fig2}
\end{figure}
The $\psi(2S)\rightarrow J/\psi \pi^+ \pi^-$ process is flavor
disconnected. In an effective theory, the tree level diagram with a
spin-0 resonance(e.g. $f_0$) of $\pi\pi$ is illustrated in
Fig.~{\ref{fig2}}. The gluon-scalar vertex $\Gamma_2^\beta$ has to
be proportional to $(k_1^\beta+k_2^\beta)$. When contracted with the
gluon propagator and the $\psi(2S)J/\psi$-gluon vertex
$\Gamma_2^{\mu\nu\alpha}$, gauge invariance will ensure the
contribution vanish. For spin-$n$($n\ge 2$) meson, the gluon-meson
vertex contracted with the spin-$n$ meson propagator will contain a
factor $(k_1+k_2)^2-m^2$, which gives a zero on the mass shell
instead of a pole. So we come to the conclusion:{\em there is no
$\pi$-$\pi$ resonance other than those with spin-1.}

Because the isospins of $\psi(2S)$ and $J/\psi$ are all zero,
isospin symmetry will require the two pions to form an
isospin singlet, which is symmetric under the interchange of
the two pions' four-momenta. Since the amplitude for an odd spin meson's decay
into two pions will be anti-symmetric when interchanging the pions' momenta,
we see that any spin-1 resonance here is suppressed by isospin symmetry.

Nor can we find any strong resonance in $J/\psi\pi^\pm$ channels.
$\psi(2S)$ decays  mainly through contact interactions.
Therefor, $M_{\psi'}$(or heavy quark mass) can be taken as the typical
energy scale of the process.

Not all terms in Eq.~(\ref{eq:p+p-}) should be kept when fitting
data. CP violating ${\cal O}(k^2)$ terms, $f_5$, $f_7$ and
$f_{11}$, are strongly suppressed and we can not to see them when
we have not enough data. $f_9$ and $f_{12}$ terms should also be
dropped because they are ${\cal O}(k^2)$ and break isospin
symmetry. Eight terms are kept:
\begin{eqnarray}
\Gamma^{\mu\nu} & = & \left\{ f_0 +\frac{[p\cdot(k_1-k_2)]}{M_{\psi'}^2}f_1
+\frac{(k_1\cdot k_2)}{M_{\psi'}^2}f_2
+\frac{[p\cdot (k_1-k_2)]^2}{M_{\psi'}^4}f_3 \right\}g^{\mu\nu}
\nonumber \\ & & +\frac{(k_1^\mu k_1^\nu +k_2^\mu k_2^\nu)}{M_{\psi'}^2}f_4 
 +\frac{(k_1^\mu k_2^\nu +k_2^\mu k_1^\nu)}{M_{\psi'}^2}f_6
+i\frac{(A_1^{\mu\nu} +A_2^{\mu\nu})}{M_{\psi'}^2}f_8
+i\frac{(A_1^{\mu\nu} -A_2^{\mu\nu})}{M_{\psi'}^2}f_{10}.
\end{eqnarray}
The $f_1$ and $f_8$ term in the amplitude are CP-violating.

\begin{table}
\caption{Symmetry properties of the amplitude. The symbol ``$\times$'' under
a parameter means  that the corresponding symmetry is violated when the 
parameter has a non-zero value.}
\begin{tabular}{cccccccccccccc}
      & $f_0$ & $f_1$ & $f_2$ & $f_3$    & $f_4$    & $f_5$    & 
$f_6$ & $f_7$ & $f_8$ & $f_9$ & $f_{10}$ & $f_{11}$ & $f_{12}$ \\ \hline
Order of $k$ & 0 & 1 & 2 & 2 & 2 & 2 & 2 & 2 & 1 & 2 & 1 & 2 & 2 \\
Isospin Symmetry &
            $\surd$  & $\times$ & $\surd$  & $\surd$  & $\surd$  & $\times$ &
            $\surd$  & $\times$ & $\surd$  & $\times$ & $\times$ & $\surd$  &
            $\times$ \\
C Parity  & $\surd$  & $\times$ & $\surd$  & $\surd$  & $\surd$  & $\times$ &
            $\surd$  & $\times$ & $\surd$  & $\times$ & $\times$ & $\surd$  &
            $\times$ \\
P Parity  & $\surd$  & $\surd$  & $\surd$  & $\surd$  & $\surd$  & $\surd$  &
            $\surd$  & $\surd$  & $\times$ & $\times$ & $\times$ & $\times$ &
            $\times$ \\
CP Parity & $\surd$  & $\times$ & $\surd$  & $\surd$  & $\surd$  & $\times$ &
            $\surd$  & $\times$ & $\times$ & $\surd$  & $\surd$  & $\times$ &
            $\surd$ 
\end{tabular}\label{tab:par}
\end{table}

For the process $\psi(2S)\rightarrow J/\psi \pi^0 \pi^0$, boson
symmetry of the two pions demands the amplitude invariant when the
four-momenta $k_1$ and $k_2$ are interchanged. Similar analysis
leads to an effective vertex of the form
\begin{eqnarray}
\Gamma'^{\mu\nu} & = & 
\left\{ g_0 +\frac{(k_1\cdot k_2)}{M_{\psi'}^2}g_1
+\frac{[p\cdot (k_1-k_2)]^2}{M_{\psi'}^4}g_2 \right\}g^{\mu\nu} 
+\frac{(k_1^\mu k_1^\nu+k_2^\mu k_2^\nu)}{M_{\psi'}^2}g_3 \nonumber \\
& & +\frac{(k_1^\mu k_2^\nu +k_2^\mu k_1^\nu)}{M_{\psi'}^2}g_4 
+i\frac{(A_1^{\mu\nu} +A_2^{\mu\nu})}{M_{\psi'}^2}g_5
+i\frac{(A_1^{\mu\nu}-A_2^{\mu\nu})[p\cdot(k_1-k_2)]}{M_{\psi'}^4}g_6,
\end{eqnarray}
and the Feynman amplitude is
\begin{equation}
{\cal
M'}_{\lambda,\sigma}(p,q,k_1,k_2)=e^\mu_{\psi'}(\vec{p},\lambda)
e^{*\nu}_{J/\psi}(\vec{q},\sigma) {\Gamma'}_{\mu\nu}.
\end{equation}
The $g_0$, $g_1$, $g_2$, $g_3$ and $g_4$ terms are CP conserving,
while $g_5$, $g_6$ terms are CP violating. Argument for why we take
$M_{\psi'}$ as the energy scale is similar to the above, and we only need 
to point out the fact that {\em any odd spin particle's decay into
two identical spin-0 particles is strictly forbidden by boson
symmetry}. Relative phases between $g_0$, $g_1$, $g_2$, $g_3$ and
$g_4$ are small. The CP breaking ${\cal O}(k^2)$ term $g_6$ can
be set to zero when one fits data since it is strongly suppressed.

The cross section of such a three-body decay depends on five variables:
$E_1$ and $E_2$, the energies of the two pions; and $(\alpha,\beta,
\gamma)$, the Euler angles that specify the orientation of the final
system relative to the initial particle~\cite{pdg98}. The cross
section of $\psi(2S)\rightarrow J/\psi\pi\pi$ does not depend on the
variable $\alpha$ provided that $\psi(2S)$ is produced by $e^+e^-$ collision,
and the $\alpha$ represents the rotation angle around the beam line.
We have
\begin{eqnarray}
& k_1\cdot k_2 = 
(E_1+E_2)M_{\psi'}-\frac{1}{2}(M_{\psi'}^2-M_{J/\psi}^2+2m_\pi^2),\\
& p\cdot (k_1-k_2) = 
(E_1-E_2)M_{\psi'}
\end{eqnarray}
and the nine independent terms in Eq.~(\ref{eq:id}) represent different angular
distributions. Those terms proportional to $|f_1|^2$,  $|f_8|^2$, $f_1f_8^*$ 
and $f_1^*f_8$ in the cross section can be ignored since it is relatively 
small. The coherent part that proportional $f_8$ (or $f_8^*$) has a 
significantly different distribution comparing with backgrounds. 
The part proportional to $f_1$ (or $f_1^*$) is an odd distribution when 
we exchange $E_1$ with $E_2$ in the Dalitz plot, 
while the backgrounds are even distributions.
These two facts will help us to distinguish the CP-violating parts from
background contributions, and will remarkably improve the precision of
the measurement of the CP-violating parameters $f_1$ and $f_8$.
Unlike the case in the decays of $K^0-\bar{K}^0$ complex, 
CP violating terms here are not mixed with backgrounds.
Fitting experimental data with such differential cross sections can
give more information on CP violation than the partial width, 
{\it e.g.}, if we integrate out the variable $p\cdot(k_1-k_2)$, the
part proportional to $f_1$(or $f_1^*$) will vanish.  

We would like to point out that to sum over the spins of $J/\psi$  or
averaging over the spins of $\psi(2S)$ in the cross section when comparing
with data(as some references) is incorrect. $\psi(2S)$ are polarized.
The density matrix of $\psi(2S)$ is determined by a vector coupling
with $e^+e^-$ at $10^{-4}$ precision. And because $J/\psi$ is re-constructed
through $e^+e^-$ and $\mu^+\mu^-$ channels, $J/\psi$ with helicity-0 has
a much less chance to be selected than those with helicity-$\pm$1. 

One might doubt if the CP-violating effects in the processes can
be measured, given that they are mediated by the strong interactions.
The Standard Model prediction of the CP-violating effects in the
processes are difficult to evaluate and it is out of the scope of
our present paper. However, we can give a rough estimation. The ratio of
the weak interactions and the strong interactions at $c\bar{c}$-meson energy
scale should be taken as $G_F m_{J/\psi}^2\sim 10^{-4}$. 
There is a unique source for CP violation in 
the Standard Model. Any CP-violating effects in the Standard Model
is proportional to the rephasing-invariant 
$Im \Delta^{(4)}$~\cite{jarlskog85}. 
The magnitude of CP violation comparing with the weak interactions can
be represented by the $\eta$-parameter measured in $K^0$ decays,
$\eta\sim 2\times 10^{-3}$~\cite{pdg98}. So we come to the conclusion
that CP-violating effect in the Standard Model, 
when compared with the strong interactions, is of the order $10^{-7}$.

\vspace{0.5cm}
\begin{figure}[htb]
\centerline{\psfig{figure=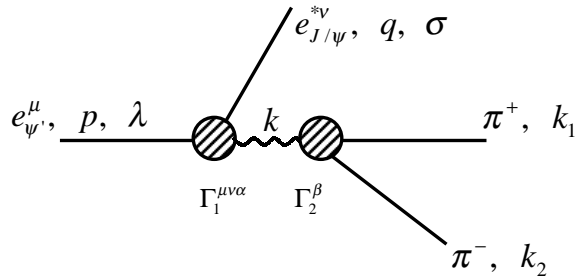,width=3in}}
\caption{Single gluon transition violates isospin symmetry.}\label{fig3}
\end{figure}

But it is not the case when a process is flavor disconnected
and the strong interactions in it are much smaller. 
As we have discussed above, 
there is no resonance in $\psi'\rightarrow J/\psi\pi\pi$ processes. 
Now suppose the decay is through single gluon exchange, 
as shown in Fig.~\ref{fig3}. Gauge invariance demands that
\begin{equation}
k^\alpha \Gamma_{1\mu\nu\alpha} e_{\psi'}^{\mu} e_{J/\psi}^{*\nu}=0,
\end{equation}
so the only possible form of the gluon-$\pi$-$\pi$ vertex is
\begin{equation}
\Gamma_2^\beta = c (k_1-k_2)^\beta,
\end{equation}
with $c$ a constant.  The two pions must form a isospin singlet 
provided the isospin symmetry conserved,
and this require $\Gamma_2^\beta$ symmetric when exchanging $k_1$ and $k_2$.
So the single gluon transition in this decay, comparing with flavor connected
processes, is suppressed by $\alpha_S$ and isospin symmetry. 
Two-gluon transitions keeping isospin symmetry are suppressed by 
$\alpha_S^2$. Be advised that $\alpha_S$ is not the coupling constant in 
QCD which is very large at low energy. Here it is the effective coupling 
of gluon and should be very small. The precise value of such transitions
are very difficult to calculate. However, we can evaluate such a suppression
by comparing the decay $\phi \rightarrow \pi^+\pi^-$ with
$\omega \rightarrow \pi^+\pi^-$.
The difference between two processes is
that the former one is flavor disconnected(both processes violate 
isospin symmetry). 
The amplitude of the two processes are all of the form
\begin{equation}
A_\lambda = c_{\phi,\omega} e^\mu(p, \lambda) (p_{\pi^+}-p_{\pi^-})_\mu,
\end{equation}
here $c_\phi$ and $c_\omega$ are coupling constants. 
The ratio of the partial width of the two 
processes is
\begin{equation}
\frac{(1-4m_\pi^2/m_\phi^2)^{3/2}|c_{\phi}|^2 m_\phi}
{(1-4m_\pi^2/m_\omega^2)^{3/2}|c_{\omega}|^2 m_\omega}
\approx 1.4 \left|\frac{c_\phi}{c_\omega}\right|^{2}
\end{equation}
The experimental value of the partial widths for the
decay $\phi \rightarrow \pi^+\pi^-$ is $4.43\times (8\times 10^{-5})
\approx 3.5\times 10^{-4}$ MeV, and $8.41\times 2.21\% \approx 0.19$ MeV
for $\omega\rightarrow \pi^+\pi^-$\cite{pdg98}. 
One can find  that the coupling constant of a flavor disconnected
vertex is suppressed by a factor of $10^{-2}$. 
Now we can conclude that the CP-violating
effect predicted by the Standard Model in the process 
$\psi(2S)\rightarrow J/\psi\pi\pi$
is of the order $10^{-5}$, comparing with the background amplitudes.
For the process $\Upsilon(2S)\rightarrow\Upsilon(1S)$, it
is estimated to be of $10^{-4}$ order.

It is believed that the Standard Model does not provide the complete
description of CP violation in nature. Almost any extension of the
Standard Model has additional sources of CP violating effects.
In addition, theories that explain the observed baryon asymmetry of 
the universe must include new sources of CP violation~\cite{cohen93}. 
The Standard Model can not generate a large enough matter-antimatter 
imbalance to produce the baryon number to entropy ratio observed in 
the universe today~\cite{farrar94,gavela94,huet95}. If the CP violating
effect beyond the Standard Model is about ten times larger, it will be
of the order $10^{-4}$ comparing with backgrounds. 

BES has accumulated $3.8\times 10^6$ $\psi(2S)$
events~\cite{bes98}. These events even make it
possible to see the $\pi^+\pi^-$ D-wave (${\cal O}(k^2)$) in the
$J/\psi \pi^+ \pi^-$ decay mode~\cite{yan99,bes99}, although the D-wave is
highly suppressed~\cite{shifman90}. Noticing that the leading CP violating
terms ($f_1$,$f_8$ and $g_5$) are ${\cal O}(k)$, it is significative to
determine whether they are zero using current data. With
BEPC has the ability to accumulate $10^8$
events within a few months, and it is scheduled for a upgrade.
The upgraded BEPC will be able to accumulate $10^9-10^{10}$ events
and can measure the CP violating parameters at a high accuracy.
Provided that the CP violating effects beyond the Standard Model are
of $10^{-4}$ order, they can be detected in the near future. 

We have given the model independent amplitudes for the decay
$\psi(2S)\rightarrow J/\psi\pi\pi$ and suggest to search for CP
violation in these processes. It is practical to measure the CP violation 
parameters. And it has the advantage that CP violation observables can be 
directly measured, so that one does not need to track two or 
more CP-conjugate processes. The extension of our results to the case of
$\Upsilon(2S)\rightarrow\Upsilon(1S)\pi\pi$ and
$\Upsilon(3S)\rightarrow\Upsilon(2S)\pi\pi$ is straightforward.
$\Upsilon(2S)$ and $\Upsilon(3S)$ can be produced at B-factories.

{\it Acknowledgement}: JJZ would like to thank Dr. Xiao-Jun Wang and 
Ting-Liang Zhuang for discussions. 
The work is partly supported by funds from IHEP and NSF of
China through C N Yang.



\end{document}